\def\ii{\'{\i}}
\def\c,{\c c}

\def\beg{\begin{equation}}
\def\fim{\end{equation}}
\def\delh{\frac{\partial h(x,t)}{\partial t}}

\documentstyle[preprint,aps]{revtex}
\begin{document}

\title{Scaling Exponents of Rough Surfaces Generated by the Domany-Kinzel
Cellular Automaton}

\author{A.P.F. Atman  \thanks{atman@fisica.ufmg.br},
	Ronald Dickman \thanks{dickman@fisica.ufmg.br},
	J.G. Moreira \thanks{jmoreira@fisica.ufmg.br}	
	} 

\address{Departamento de F\ii sica, Instituto de Ci\^encias Exatas,\\
Universidade Federal de Minas Gerais, C. P. 702\\ 
30123-970, Belo Horizonte, MG - Brazil}
\date{\today}
\maketitle
  
\begin{abstract}
The critical behavior at the frozen/active transition in the Domany-Kinzel 
stochastic cellular automaton (DKCA) is studied {\it via} a surface growth 
process in (1+1) dimensions. At criticality, this process presents a kinetic 
roughening transition; we measure the critical exponents in simulations. 
Two update schemes are considered: 
in the symmetric scheme, the growth surfaces belong to the Directed Percolation 
(DP) universality class, except at one terminal point. At this point, the phase 
transition is discontinuous and the surfaces belong to the Compact Directed 
Percolation universality class. The relabeling of space-time points in the
nonsymmetric scheme alters the surface growth dramatically.
The critical behavior of rough surfaces at 
the nonchaotic/chaotic transition is also studied using the damage spreading 
technique; the exponents confirm DP values for the symmetric scheme. 


\end{abstract}
\newpage

\section{Introduction}
The one-dimensional Domany-Kinzel stochastic cellular automaton (DKCA) is a 
completely discrete sys\-tem - temporally, spatially and in its state space - 
with applications in physics, chemistry, biology, computer
science, etc. \cite{dk,wolfram}. The DKCA also attracts interest as a particle
system affording a test of certain conjectures regarding nonequilibrium critical
phenomena \cite{kinzel}. 
The DKCA has a unique absorbing (``vacuum") state; its phase diagram presents a 
critical line separating this absorbing phase from an active phase.
Models with one absorbing state have been conjectured to belong generically 
to the directed percolation (DP) universality class \cite{grass1}. 
There is also good numerical evidence \cite{kinzel} 
that the critical behavior along the transition line 
in the DKCA belongs to the DP class, except at one of the terminal points, 
where the asymptotic behavior is known exactly and belongs to the compact 
directed percolation (CDP) universality class \cite{essam,dicktret}. At this 
terminal 
point the transition is discontinuous and we have in fact two absorbing states: 
the vacuum and the completely filled state. Using the damage spreading 
technique, Martins {\it et al.} \cite{martins} found another critical line 
separating the active phase into a nonchaotic and a chaotic phase. There is 
numerical evidence that the critical 
behavior along this transition line also belongs to the DP class, as 
expected on the basis of universality \cite{grass2}.

The surface growth process generated by cellular automata (CA) was 
proposed by de Salles {\it et al.} \cite{sales1} to study
Wolfram's deterministic CA. These authors also used this process to
identify the frozen/active phase transition in the DKCA \cite{sales2}, 
where they showed that the Hurst exponent $H$ attains a maximum at the 
transition. Recently, Atman and Moreira \cite{atman} demonstrated that the 
growth exponent $\beta_w$ also presents a cusp at criticality, and is more 
appropriate for detecting phase transitions than the method of de Salles 
{\it et al.}. They used it to construct the  
DKCA phase diagram and conjectured that the growth exponent method also can be
used to detect phase transitions in other kinds of models. Recently, Redinz and
Martins \cite{reid} used the Hurst exponent method to find first and second 
order phase transitions
in the $q$-state Potts model (for $q = 1, 3, 5$ and $10$). 
Bhattacharyya \cite{bhatta2} studied the dynamic critical 
properties of a related one-dimensional probabilistic cellular automaton, using 
two different procedures to generate the surface growth process. One of them 
is identical to the growth process studied by de Salles {\it et al.}, 
and belongs to DP universality class. Thus, the conjectures presented by 
Bhattacharyya \cite{bhatta2} which imply DP-like scaling for this model are 
also valid here.

Basically, the procedure used by de Salles {\it et al.}\cite{sales2} and by 
Bhattacharyya \cite{bhatta2},
consists in transforming the spatiotemporal patterns 
generated by the DKCA to a solid-on-solid (SOS) particle deposition.
At the critical line, it yields a surface growth process with 
kinetic roughening, and the critical exponents can be measured following the 
scaling concepts developed by Family and Vics\'ek \cite{family}. Very 
recently, similar methods were used by Lauritsen and Alava \cite{lauri} 
to study the Edwards-Wilkinson equation with columnar noise, 
by Vespignani {\it et al.} \cite{vesp} to study sandpile models,
and by Dickman and 
Mu\~noz \cite{dickman}, to study the contact process (CP). 

In this work, we measure the scaling exponents at criticality
in simulations, and compare them with known DP and CDP values. We use 
two different schemes to update the automaton, which lead 
to entirely different surface growth scaling properties. 
In Section II, we define 
the DKCA, describe the two update schemes, and
show how the surface growth process is generated, at the
frozen/active and nonchaotic/chaotic transitions. In Section III, we present 
our numerical results and discuss the values obtained, comparing them
with the predictions for the scaling exponents proposed by
Bhattacharyya \cite{bhatta2} and Dickman and Mu\~noz \cite{dickman}. We discuss 
our conclusions in Section IV.

\section{DKCA Surface Growth Process}

\subsection{Model}
 
The DKCA was proposed by Domany and Kinzel \cite{dk}, who showed 
the existence of two phases: {\it active} and {\it frozen}. A more detailed 
study, using simulation, was performed by Martins {\it et al.} 
\cite{martins}, in which a new phase within the active region - a 
{\it chaotic} phase - 
was discovered through the damage spreading technique. 

The DKCA consists of a linear chain of $L$ sites ($i=1,2,...,L$), with 
periodic boundaries, where each site $i$ has two possible states, 
conveniently denoted by 
$\sigma_i = 0,1$. The state of the system at time $t$ is given
by the set \{$\sigma_{i}(t)$\}. In contrast to the deterministic CA studied by 
Wolfram \cite{wolfram}, the DKCA is probabilistic: the rules for 
updating the system are given by conditional probabilities,
which depend on the neighbors. We study two different schemes, one symmetric, 
the other nonsymmetric. The symmetric scheme is the original one
proposed by Domany-Kinzel \cite{dk}, while the nonsymmetric was used \cite{nagy}
to simplify the algorithm; its authors affirm that the two schemes
are topologically equivalent. 

In the symmetric scheme, the process $\sigma_i(t)$ is defined on space-time
points with $i\!+\!t$ {\it even}. The state of site $i$ at time $t\!+\!1$ 
depends on $\sigma_{i\!-\!1}(t)$ and $\sigma_{i\!+\!1}(t)$ via the transition 
probability \\
$P[ \sigma_{i}(t\!+\!1) | \sigma_{i\!-\!1}(t),\sigma_{i\!+\!1}(t)]$,
which takes the form: \\
$P(1|0,1) = P(1|1,0) = p_1~,$ \\
$P(1|1,1) = p_2~,$ \\
$P(1|0,0) = 0~.$ \\
Evidently, 
$P(0| \sigma_{i-1}, \sigma_{i+1} ) = 1 - P(1| \sigma_{i-1}, \sigma_{i+1})$.

In the non\-sym\-me\-tric scheme, the process is defined on {\it all} 
space-time points. The state of site $i$ at time $t\!+\!1$
depends on $\sigma_{i\!-\!1} (t)$ and $\sigma_{i} (t)$, rather than on 
$\sigma_{i-1} (t)$ and $\sigma_{i+1} (t)$, as in the symmetric scheme.
The transition probability
$P[\sigma_i(t\!+\!1)|\sigma_{i\!-\!1} (t),\sigma_{i} (t)]$
is identical to 
$P[\sigma_i(t\!+\!1)|\sigma_{i\!-\!1} (t),\sigma_{i\!+\!1} (t)]$
given above for the symmetric case.

It is easy to see that the two schemes are connected via a simple 
relabelling of space-time points (see Fig. 1).
Consider, for example, a {\it history} $h^S$ in the symmetric scheme,
that is, a sequence of configurations $\{\sigma_i^S (t)\}$ for $t=0,...,T$
($T$ finite), starting at $t=0$ with a finite number of active
sites. Let the probability of this history, given the initial
configuration, be $P[h^S|\{\sigma^S(0)\}]$.  
A history $h^{NS}$ in the nonsymmetric scheme can be defined in
the same manner. Note that there is a one-to-one correspondence between
histories in the two schemes, given by 
\begin{equation}
\sigma_i^{NS} (t) \equiv \sigma_{2i-t}^S (t)  \;.
\label{map}
\end{equation}
Since the transition probabilities in the two schemes are identical,
the probabilities of corresponding histories are as well.
To extend this correspondence to systems with periodic boundaries,
we note that if the nonsymmetric system has $L$ sites, then the
corresponding symmetric one has $2L$ sites; in the mapping defined above,
we now take $i^{S} = 2i^{NS}\! -\! t \;\;\; (\mbox{mod} \; 2L)$.

An immediate result of this correspondence is that all scaling
properties (e.g., critical exponents), as well as nonuniversal properties 
(e.g., phase boundaries between frozen, active, and chaotic 
phases in the $p_1 - p_2$ plane), are identical in the two
schemes.  Corresponding histories naturally look different in the
two schemes:  the nonsymmetric scheme represents a rotating frame of
reference in which, moreover, distances are rescaled by a factor of
1/2.  Thus, for $p_1=1/2$ and $p_2=1$, an interface between domains of
1's and 0's executes an unbiased random walk in the symmetric scheme,
while in the nonsymmetric case such an interface has a mean 
velocity of 1/2.  The ``light cone" $i = \pm t$ in the symmetric scheme
becomes the pair of lines $i=0$ and $i=t$ in the nonsymmetric case.
As will be seen below, this difference in frames of 
reference has important consequences for the surface dynamics in the
nonsymmetric scheme.

Depending on the values of the parameters ($p_1,p_2$), the asymptotic 
($t \rightarrow \infty$) state of the system is either {\it frozen}, 
with all sites having value 0, or has a finite fraction of sites with value 1, 
the {\it active} state. This is a second order phase transition, characterized 
by the critical exponents ofthe DP universality class.

\subsection{Interface Representation}

The surface growth process consists in accumulating 
(summing) all the values assumed by the variables $\sigma_{i}(\tau)$ over 
the first $t$ time steps:
\beg\label{soma}
h_{i}(t) \equiv \sum^{t}_{\tau=0} \sigma_{i}(\tau)~~.
\fim
 
The differences between the schemes become explicit at this point. In Figure 2
we show the temporal evolution of the automaton and the profiles 
generated by the accumulation method, close to criticality 
$(p_2=0.5, p_1=0.75)$, in each scheme. It is evident that the two
schemes lead to entirely different profiles. (In this figure, we choose an 
initial condition of a single active site, to highlight the 
evolution of the automaton and profiles.) 
 
Thus, we obtain growth processes, the nature of whose correlations can be 
investigated through the analysis of the roughness $w(L,t)$ \cite{barab},
defined by
\beg\label{rugos}
w^{2}(L,t) = \frac{1}{L} \left< \sum^{L}_{i=1} \left( h_{i}(t) - \overline{h}(t) 
\right)^{2} \right>~~,
\fim
\noindent where $\overline{h}(t)$ is the mean value of $h_{i}(t)$ at time $t$,
and the brackets $<\ldots>$ denote an average over realizations.

We expect that $w(L,t)$ has the scaling form \cite{family}
\beg\label{scal}
w(L,t) \sim L^{\alpha} f \left( \frac{t}{L^{z}} \right)~~,
\fim
\noindent where $f(u)$ is a universal scaling function, $\alpha$ is the 
roughness exponent, $z = \alpha/ \beta_w$ is the dynamic exponent and 
$\beta_w$ is the growth exponent. The function $f(u) = constant$, at large 
times ($t \gg L^{z}$), and $f(u) \sim u^{\beta_w}$ at short times 
($t \ll L^{z}$).
So at short times, we expect $w(t) \sim t^{\beta_w}$; we  
measure $\beta_w$ from the slope of the $\log - \log$ plot of 
$w(L,t)$ versus $t$. At large times, the roughness saturates and 
becomes $L$-dependent: $w(L,\infty) \sim L^{\alpha}$. The crossover time,
$t_{\times}$, between these two regimes grows as $t_{\times} \sim L^z$. 
The exponents $\alpha$ and $z$ are defined in the frozen phase just at the 
transition line. In the active phase, the roughness does 
not saturate, growing instead as  $w(L,t) \sim t^{1/2}$, corresponding to 
uncorrelated growth \cite{atman}. The relations above, used to measure 
the scaling exponents, are valid just at criticality.

The profiles have self-affine properties quantified by the Hurst 
exponent $H$, defined {\it via}
\beg
W(\epsilon) \sim \epsilon^H~,
\fim
\noindent where $W(\epsilon)$ is the width of the interface on length scale
$\epsilon$. We measure the Hurst exponent, $H$, in a profile generated very 
close to the transition. We apply the method introduced by Moreira {\it et al} 
\cite{moreira}, that consists in measuring the roughness around 
the straight line determined by a least-squares fit to a segment of the
profile. The roughness $W(L,\epsilon,t)$, 
at the scale $\epsilon$, is given by
\beg
W(L,\epsilon,t) = \frac{1}{L} \sum_{i=1}^{L} w_i(\epsilon,t)
\fim
\noindent where the local roughness $w_i(\epsilon,t)$ is defined by
\[
w_i^2(\epsilon,t)=\frac{1}{2\epsilon +1} \sum_{j=i-\epsilon}^{j=i+\epsilon}
\{h_j(t) - [a_i(\epsilon)x_j + b_j(\epsilon)]\}^2~.
\]
\noindent $a_i(\epsilon)$ and $b_i(\epsilon)$ are the linear 
fitting parameters to the profile on the interval 
$[i-\epsilon, i+\epsilon]$ centered at site $i$. 

\subsection{Damage Spreading}

Martins {\it et al.} \cite{martins} used the damage spreading technique to show 
that the active phase of the DKCA in fact consists of two phases, chaotic and 
nonchaotic. The order parameter of this transition is the difference between 
two replicas started with different initial configurations. One lets 
the system 
evolve until it attains a stationary state, and then a replica of the 
configuration 
is created with some sites altered (damage). The two replicas, one with 
state $\sigma_{i}(t)$ and the other with state $\varrho_{i}(t)$, evolve with 
the same sequence of random numbers, and the difference between the configurations  
\[
\Gamma_{i}(t) = | \sigma_{i}(t) - \varrho_{i}(t) |~,
\]
is measured. The  fraction of sites in the two replicas with 
$\sigma_i \neq \varrho_i$ is called the Hamming distance, defined as 
\[
D_H (t) = \frac{1}{L}\sum_i \Gamma_i (t)~.
\]
The stationary Hamming distance is null in the non-chaotic phase and positive 
in the chaotic phase. 

To study the chaotic/non-chaotic boundary, we use a slightly different method, 
where the difference between the two automata is used to generate the surface
growth process, as we did in the accumulation method
\beg\label{acc2}
h_{i}(t) = \sum^{t}_{\tau=0} \Gamma_{i}(\tau)~.
\fim
Thus, the profile generated by the difference between the replicas behaves 
exactly as the profiles generated in the frozen/active boundary: the roughness 
reaches a stationary value in the non-chaotic phase and grows indefinitely in 
the chaotic phase. This behavior can be understood if we note that the 
difference between the replicas vanishes in the non-chaotic phase, 
which implies no contribution to the height $h_{i}(t)$, and is positive in 
chaotic phase, implying steady growth in the height. 

In order to preserve the stationary 
density of active sites we generate a ``rotation'' 
damage at a certain time $t_0$, in which
the replica is rotated 180$^0$ with respect to the original system, that is
$\varrho (i, t_0) = \sigma(i+L/2, t_0)$, subject to the periodic boundary 
condition. 

\subsection{Theoretical Descriptions}

Theoretical descriptions 
of surface growth scaling at absorbing-state phase transitions
were proposed by Bhattacharyya \cite{bhatta2}, and
Dickman and Mu\~noz \cite{dickman}. 
Bhattacharrya proposed an analytical treatment in analogy to the random 
deposition (RD) process: considering an initial disordered state, the growth 
process can be described by a continuum equation, very similar to the 
RD process:
\beg\label{rd}
\delh = F + \eta(\vec{x},t)~~~~~~~~~~~~~~~~~~~~~~~~[RD]
\fim 
\noindent where $F$ is the average number of deposited particles and 
$\eta(\vec{x},t)$ corresponds to the white noise ($<\eta(\vec{x},t)>=0$) in 
the deposition. 

The difference between the surface growth processes generated by DKCA and 
RD lies in the noise correlations.
While RD involves spatially and temporally uncorrelated noise, the 
correlations in time and space developing in the DKCA appears in the noise
fluctuations of the accumulation method. 
For values of ($p1,p2$) away from the critical line (in the active phase), the 
correlation length $\xi$ and correlation time $\tau$ of the DKCA are finite, 
which means that the noise in the deposition process is correlated over short 
ranges. In this limit the noise autocorrelation decays exponentially
\cite{barab}:
\beg
<\eta(\vec{x},t),\eta(\vec{x'},t')> \sim e^{-|\vec{x}-\vec{x'}|/\xi} e^{-|t-t'|/\tau}~.
\fim

Thus, the noise appears uncorrelated for times greater than $\tau$, and the RD
exponents are obtained in this limit. This behavior was confirmed in earlier 
simulations \cite{atman}.
As we approach the transition line, $\xi$ and $\tau$ increase, and it takes 
longer for the growth process to reach the RD limit. Finally, at the critical 
line, both $\xi$ and $\tau$ diverge, and the correlations are long ranged, 
represented by a power law decay of the noise autocorrelation \cite{barab}
\beg\label{noise}
<\eta(\vec{x},t),\eta(\vec{x'},t')> \sim |\vec{x}-\vec{x'}|^{-2\beta/\nu_{\perp}} 
|t-t'|^{-2\beta/\nu_{\parallel}}~,
\fim
\noindent where $\beta$, $\nu_{\perp}$ and $\nu_{\parallel}$ are,
respectively, the critical 
exponents for the order parameter, correlation length and correlation time of 
the DKCA. 

The value of the growth exponent $\beta_w$ at the critical line can 
be derived from the continuum equation (\ref{rd}) and the noise (\ref{noise}); 
the width for an infinite substrate is expected to increase as the 
following power of the time \cite{bhatta2}
\beg
w(\infty, t) \sim t^{1-\beta/\nu_{\parallel}}~.
\fim 
As it is believed that this transition belongs to the DP universality class, the 
value of growth exponent is expected to be \cite{jensen} 
\[
\beta_w = 1- \frac{\beta}{\nu_{\parallel}} = 1-\frac{0.27649}{1.733825} \simeq 0.8405~.
\]

In previous work \cite{atman}, Atman and Moreira showed that $\beta_w$ attains 
a maximum
at the phase transition (see Fig.3), and measured its value along the 
transition line 
of the DKCA. This behavior of the exponent $\beta_w$ in the vicinity of the
phase transition can be understood as follows. The growth rate, 
$dh_i/dt$, at site $i$ is proportional (in the frame of reference 
moving with the average velocity $d<h>/dt$) to the 
{\it excess} activity  
at that site. Away from the critical point, the activity has a finite 
correlation lenght $\xi$ and correlation time $\tau$. Thus on scales much 
greater than $\xi$, $\tau$, the noise driving the surface growth is uncorrelated,
and  this process falls in the RD class, with $\beta_w = 1/2$. At the 
critical 
point, by contrast, $\xi$ and $\tau$ diverge and we have instead the scaling 
relation $\beta_w = 1 -\theta$ \cite{dickman}, where the exponent $\theta$ is
defined through the relation $\rho (t) \sim t^{-\theta}$ for the initial decay
of the activity density $\rho$, at the critical point, starting from 
$\rho (0) = 1$.
Since, in one dimension, $1-\theta > 1/2$,
we expect a jump in $\beta_w$ at the phase boundary. In simulations of 
finite-sized systems, we expect not a discontinuity in $\beta_w$ but a sharp 
peak at the transition - see Figure 3 (very near to the critical point, 
$\xi > L$, so that independently fluctuating regions are not present in the 
simulation. Below the transition, the apparent value of $\beta_w \rightarrow 0$
due to the short lifetime of the activity). It is interesting note that for the 
DP universality class, $1-\theta \simeq  0.55, 0.27$ and 0 for $d = 2,3$ and 4, 
respectively. Thus we should expect $\beta_w$ to {\it decrease} at the phase boundary
in $d=3$.

Since the saturation of surface width is forced by the DKCA, the 
crossover time $t_\times$ behaves exactly as in DP, and the dynamic 
exponent is given by:
\beg 
t_\times \sim L^z~,~~~z=z^{DP} = \nu_{\parallel}/\nu_{\perp} \simeq 1.5808~.
\fim
Thus, the roughness exponent at the criticality is given by 
\beg 
\alpha = z \cdot \beta_w \simeq 1.3286~.
\fim

Dickman and Mu\~noz \cite{dickman} studied the contact process using the surface
growth representation. They demonstrated that the Hurst exponent shows 
clear signs of anomalous scaling ($\alpha > H$), but no evidence of multiscaling.
They verified L\'opez' 
scaling relation \cite{lopez}: $H = \alpha - z\kappa$. Here, $\kappa$ is the exponent
associated with the divergence of the mean-square height gradient in the 
continuum growth equation that describes the contact process and 
related models, such as DP. Inserting the known values in the relation above
we have for the Hurst exponent in the DP class:
\beg
H=\alpha - z \kappa \simeq 0.643~.
\fim

\section{Results}

In Table 1, we summarize our results for the scaling exponents at the 
frozen/active and nonchaotic/chaotic transitions, 
and compare
them with the values for the DP and CPD universality classes. 

To extract the exponent values from our simulation data, we used the relations
$w(L, \infty) \sim L^{\alpha}$, valid at large times, $w(L,t) \sim t^{\beta_w}$,
valid at short times and $t_{\times}(L) \sim L^{z}$. The results show a
strong dependence on the scheme used - symmetric or nonsymmetric. In the 
simulations, we average 10000, 5000, 2500, 1000, 500, 250 and 100
samples at the critical point ($p_2^c, p_1^c$) in systems with $L=$ 50, 100, 
200, 500, 1000, 2000 and 5000 sites, respectively. The initial condition 
in these samples was random, with 50\% of sites active.

The Hurst exponent was measured following the procedure explained in 
subsection 2.2.
The results, shown in Table I, represent an average over 100 random
initial configurations in a system with $L =10000$. We observe a 
significant change in the value of the Hurst exponent depending on the scheme 
used to update the automaton: in the symmetric scheme, we have $H > 1/2$, 
denoting a
positive correlation in the profile; in the nonsymmetric scheme $H=0.25(3)$,
denoting a negative correlation. This behavior can be understood
considering the nonsymmetric scheme as a deposition over a moving reference 
frame, which implies a lateral propagation of correlations. 

\subsection{Symmetric scheme}

For the symmetric scheme, our results for the critical exponents agree with the
DP values, except at terminal point $p_2 = 1$ where CDP values were obtained. 
The critical points ($p_2^c, p_1^c$) were determined through the growth 
exponent method 
\cite{atman} (Figure 3). This method consists in fixing $p_2$ and varying $p_1$ 
until the maximum of the growth exponent $\beta_w$ is attained.  

To determine the crossover time $t_{\times}(L)$,
we plot the fraction of realizations with at least one active site as a 
function of time (see Fig. 4), and define the crossover time such that half 
of the initial sample has frozen. The inset of
Figure 4 shows the power law behavior of the crossover time; the slope
of this line corresponds to the exponent $z$.

To obtain the saturation width, we let all samples evolve to the absorbing state,
at a given system size, and determine the final averaged roughness. The exponent
$\alpha$ was measured as the slope of $w(L, \infty)$ versus $ L$ in a log-log 
plot. In Figure 5 we present the results for the saturation roughness 
in the cases: directed percolation ($p_2=0.5$, $p_1=0.749$, symmetric), 
directed percolation ($p_2 = 0, p_1 = 0.8095$, damage spreading), compact 
directed percolation ($p_2=1$, $p_1=0.5$, symmetric) and nonsymmetric DP 
($p_2=0.5$, $p_1=0.749$, nonsymmetric).

In order to verify Family-Vics\'ek scaling, we use the measured values for the 
scaling exponents to collapse the width curves at different system sizes to a 
single curve, as shown in Figure 6. Note the collapse of the width curves, 
corroborating the Family-Vic\'ek scaling relation.  

It is important to note that the exponents $\beta_w$ and $z$ measured 
for the chaotic/nonchaotic transition at 
$p_2=0$ are 
slightly different from the exponents measured away from this point. 
This represents evidence of long-range correlations
due the coincidence of the damage spreading and 
frozen/active transitions at this point, as pointed out by Grassberger 
\cite{grass2}. 

\subsection{Nonsymmetric scheme}

A significant change in the roughening occurs in the nonsymmetric scheme, as 
shown in Figure 7; we observe two distinct regimes in the roughness growth:
a strongly correlated regime, for times $t \stackrel < \sim L$, and a weak 
correlation regime, for longer times.
Again, this behavior can be understood by considering the nonsymmetric scheme as
deposition in a moving reference frame.  The correlations inherent in the 
dynamics are
propagated by the moving reference frame until they reach the system size; then the 
correlations due the local rules of the automaton take over, decreasing the
growth rate of the roughness. 
These two regimes
for the roughness growth implies that is not possible colapse all the curves 
using the Family-Vicsek scaling law.

The apparent exponent values measured in the nonsymmetric scheme are markedly 
smaller: $\alpha \sim 0.93(2)$ for nonchaotic/chaotic and
frozen/active (at $p_2 \neq 1$) transitions; at 
$p_2 =1, p_1=0.5$, $\alpha \sim 0.984(7)$.
As discussed above, the growth exponent presents two values, depending on the
roughness growth regime.
The dynamic exponent $z$ must assume the DP value, as discussed
in Sec. 2.1; in fact, 
$z \sim 1.6(1)$ for
nonchaotic/chaotic and frozen/active (at $p_2 \neq 1$) transitions,
while $z \sim 1.9(1)$ at $p_2=1, p_1=0.5$.

\section{Conclusions}

Growth surfaces generated by the spatiotemporal patterns
of the DKCA along its critical lines are studied. The critical roughening
exponents, expected to belong to the DP universality class, were measured
using power law relations valid at criticality. Except for the
terminal point $p_2=1$, all the scaling exponents
agree with the DP values, in the symmetric scheme, and the scaling law 
$\beta_w=\alpha/z$ remains valid. At $p_2=1$, we confirm CDP values for
the exponents. Since the fluctuations in uncorrelated regions 
are effectively superposed, it is not surprising that the apparent values
of $\beta_w$ and $\alpha$ are smaller in the nonsymmetric scheme. 
At the nonchaotic/chaotic transition, the exponents measured also agree
with the DP values.

\noindent {\bf Acknowledgements}
We thank the Brazilian agencies CNPq and Fapemig for financial support of 
this work. 



\newpage

\begin{table*}[t]
\caption{Summary of the scaling exponents values, d=1.}
\label{table:1}
\newcommand{\m}{\hphantom{$-$}}
\newcommand{\cc}[1]{\multicolumn{1}{c}{#1}}
\renewcommand{\tabcolsep}{1.5pc} 
\renewcommand{\arraystretch}{1.2} 
\begin{tabular}{@{}lllll}
\hline
Previous work		&\cc{$\alpha$}&\cc{$\beta_w$}&\cc{$z$}&\cc{$H$} \\
\hline
DP 	 		& 1.3286 & 0.8405 & 1.5808 & 0.643 \\
CP (Simulational) \cite{dickman} & 1.33 & 0.839(1) & - & 0.63(3) \\
CA (Simulational) \cite{bhatta2} & - & 0.837(11) & - & - \\
CDP 	 		& 2 & 1 & 2 & 1 \\
\hline
Present work - Symmetric scheme      	 \\
\hline
frozen/active $p_2=0.5$  	& 1.32(1) & 0.82(2) & 1.59(1) & 0.61(3) \\
frozen/active $p_2=1$ 	& 2.01(1) & 0.99(1) & 2.08(5) & 0.99(2) \\
nonchaotic/chaotic 	$p_1=1$	& 1.325(9)& 0.81(1) & 1.61(1) & 0.60(3) \\
nonchaotic/chaotic 	$p_2=0$	& 1.32(1) & 0.78(2) & 1.64(2) & 0.61(3) \\
\hline
\end{tabular}
\end{table*}

\newpage

{\large \bf Figure Captions}

\noindent {\bf Figure 1}
Spatial representation of DKCA, in symmetric (left) and nonsymmetric 
(right) schemes,
showing that the spatio-temporal patterns are identical in the two schemes, 
{\it i.e.}, corresponding histories are identical.

\noindent{\bf Figure 2}
Growth surfaces generated by spatiotemporal patterns of the DKCA with
different update schemes: the symmetric scheme is shown on the left and the 
nonsymmetric on the right. Upper panels: temporal evolution of DKCA. Black sites
are active; time increases upward. Lower: profiles 
generated by the accumulation method; the fill color is changed every 50 steps.
System size is $L=500$ for the symmetric and $L=250$ for the nonsymmetric; 
900 time steps are shown. Both systems are very close to 
criticality ($p_2=0.5$, $p_1=0.75$) in the active phase.

\noindent{\bf Figure 3}
Growth exponent $\beta_w$ in the DKCA for several system sizes in the 
symmetric scheme. Note that $\beta_w$ 
attains a maximum at the frozen/active transition and depends strongly on the 
system size. The transition point was chosen as the $\beta_w$
value at system size $L=10000$, where we observe a sharp transition. In this 
example, $p_2=0.5$ and $p_1=0.749$.

\noindent{\bf Figure 4}
Dynamic exponent $z$ for the DKCA interface representation.
The density $\rho(t)$ of active samples in function of time for several system
sizes is shown, for the symmetric scheme ($p_2 = 0.5, p_1 = 0.749$). The 
horizontal line highlight the value $\rho =1/2$, which
corresponds to the crossover time. The inset shows the crossover time 
$t_{\times}$ in function of the system sizes. The slope of this curve is the
value of the dynamical exponent $z$. The error bars are 
calculated considering an error of 1\% in the number of samples froze at a 
given time.

\noindent{\bf Figure 5}
Roughness exponent $\alpha$ for the DKCA interface representation.
Four cases are shown: directed percolation 
($p_2=0.5$, $p_1=0.749$ symmetric), directed percolation ($p_2=0$, $p_1=0.8095$ 
damage spreading),compact directed percolation 
($p_2=1$, $p_1=0.5$ symmetric) and nonsymmetric DP ($p_2=0.5$, $p_1=0.749$ 
nonsymmetric). The line is the power law regression for the data and furnishes 
the value of the roughness exponent $\alpha$. The error bars are the standard 
deviations of the saturation width over realizations at each 
system size.

\noindent{\bf Figure 6}
Family-Vics\'ek scaling. Upper panel: the width of the generated 
profiles at different
system sizes. Lower: collapse of the curves above using the exponent values
measured through numerical simulation at the criticality ($p_2=0.5$, 
$p_1=0.749$).

\noindent{\bf Figure 7}
Profile roughness behavior for nonsymmetric scheme. Note the 
two regimes in the $w(L,t) \times t$ curve; for $t \stackrel < \sim L$ we
have a srongly correlated regime with $\beta_w \sim 0.65$ and for 
$L < t < t_{\times}$, a weakly correlated regime with $\beta_w \sim 0.45$. 


\newpage

\begin{thebibliography}{99}

\bibitem{dk}
E. Domany and W. Kinzel,
\newblock {\it Phys. Rev. Lett.} {\bf 53}, 447 (1984).

\bibitem{wolfram}
S. Wolfram,
\newblock {\it Theory and Applications of Cellular Automata},
\newblock (Word Scientific, Singapore, 1986).

\bibitem{kinzel}
W. Kinzel, 
\newblock {\it Z. Physik B} {\bf 58}, 229-244 (1985).

\bibitem{grass1}
P. Grassberger,
\newblock {\it Z. Phys. B} {\bf 47}, 365 (1982);\\
H. K. Janssen, 
\newblock {\it Z. Phys. B} {\bf 42}, 151 (1981).

\bibitem{essam}
J.W. Essam,
\newblock{\it J. Phys. A} {\bf 22}, 4927 (1989). 

\bibitem{dicktret}
R. Dickman and A. Y. Tretyakov,
\newblock {\it Phys. Rev. E} {\bf 52}, 3218 (1995).

\bibitem{martins}
M. L. Martins, H. F. Verona de Resende, C. Tsallis and A. C. N. de Magalh\~aes,
\newblock {\it Phys. Rev. Lett.} {\bf 66}, 2045 (1991).

\bibitem{grass2}
P. Grassberger,
\newblock {\it J. Stat. Phys.} {\bf 79}, 13 (1995).

\bibitem{sales1}
J. A. de Sales, M. L. Martins and J. G. Moreira,
\newblock {\it Physica A} {\bf 245}, 461 (1997).

\bibitem{sales2}
J. A. de Sales, M. L. Martins and J. G. Moreira,
\newblock {\it J. Phys. A} {\bf 32}, 885 (1999).

\bibitem{atman}
A. P. F. Atman and J. G. Moreira,
\newblock {\it Eur. Phys. J. B} {\bf 16}, 501 (2000). 

\bibitem{reid}
J. A. Redinz and M. L. Martins,
\newblock Phys. Rev. E {\bf 63}, 66133 (2001).

\bibitem{bhatta2}
P. Bhattacharyya,
\newblock {\it Int. J. Mod. Phys. C} {\bf 10}, 165 (1999).

\bibitem{family}
F. Family and T. Vics\'ek,
\newblock {\it J. Phys A: Math Gen.} {\bf 18}, L75 (1985).

\bibitem{lauri}
K. B. Lauritsen and M. Alava,
\newblock e-print: cond-mat/9903346.

\bibitem{vesp}
A. Vespignani, R. Dickman, M. A. Mu\~noz, and S. Zapperi,
\newblock{\it Phys. Rev. E} {\bf 62}, 4564 (2000).

\bibitem{dickman}
R. Dickman and M. A. Mu\~noz,
\newblock{\it Phys. Rev. E} {\bf 62}, 7632 (2000).

\bibitem{nagy}
T.F. Nagy, S.D. Mahanti and C. Tsallis,
\newblock {\it Physica A} {\bf 250}, 345 (1998).

\bibitem{barab}
A.-L. Barab\'asi e H. E. Stanley, 
\newblock {\it Fractal Concepts in Surface Growth},
\newblock (Cambridge Univ. Press, Cambridge, 1995).

\bibitem{moreira}
J. G. Moreira, J. Kamphorst Leal da Silva and S. Oliffson Kamphorst,
\newblock {\it J. Phys A: Math. Gen.} {\bf 27}, 8079 (1994).

\bibitem{jensen}
I. Jensen,
\newblock {\it J. Phys A: Math. Gen.} {\bf 29}, 7013 (1996).

\bibitem{lopez}
J. M. L\'opez,
\newblock {\it Phys. Rev. Lett.} {\bf 83}, 4594 (1999).

\end{thebibliography}
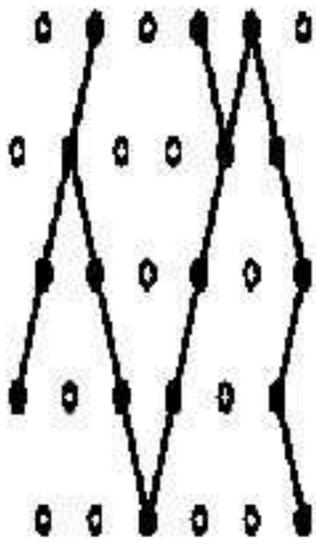
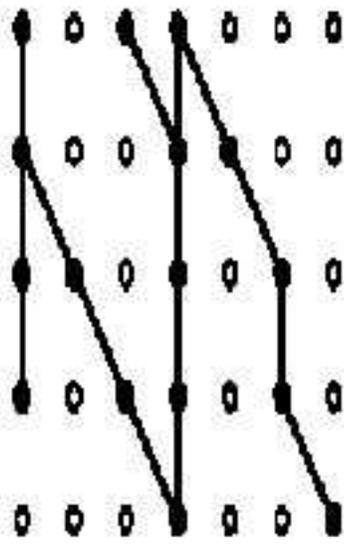
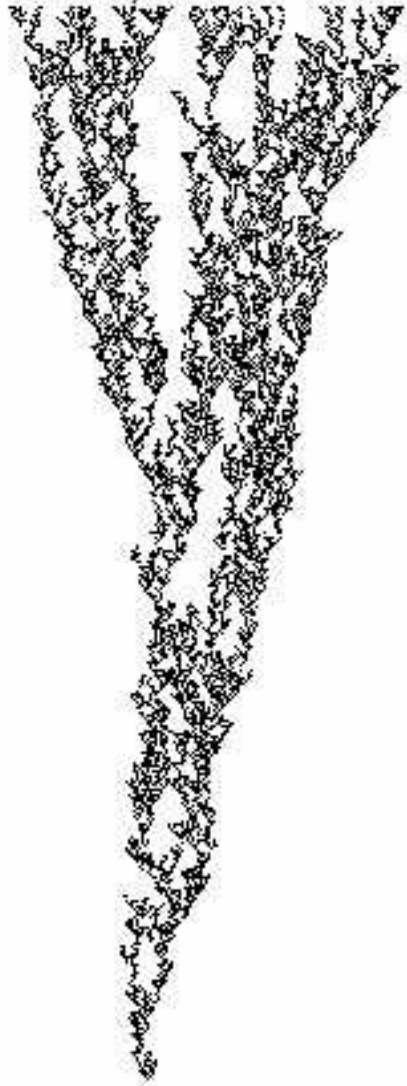
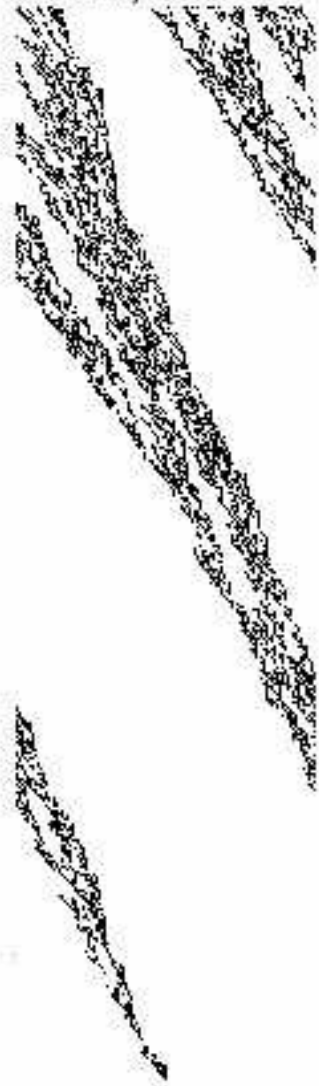
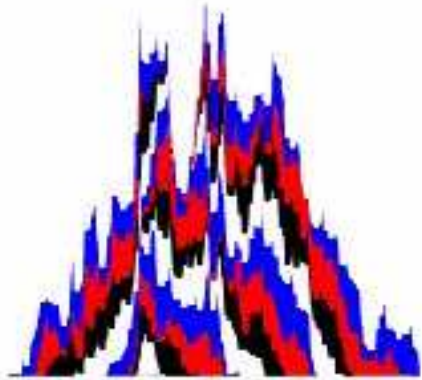
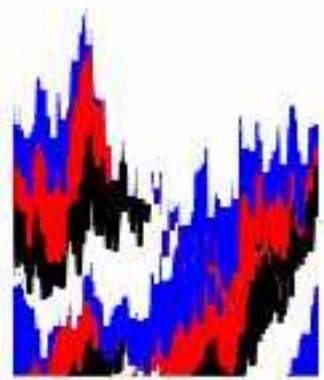
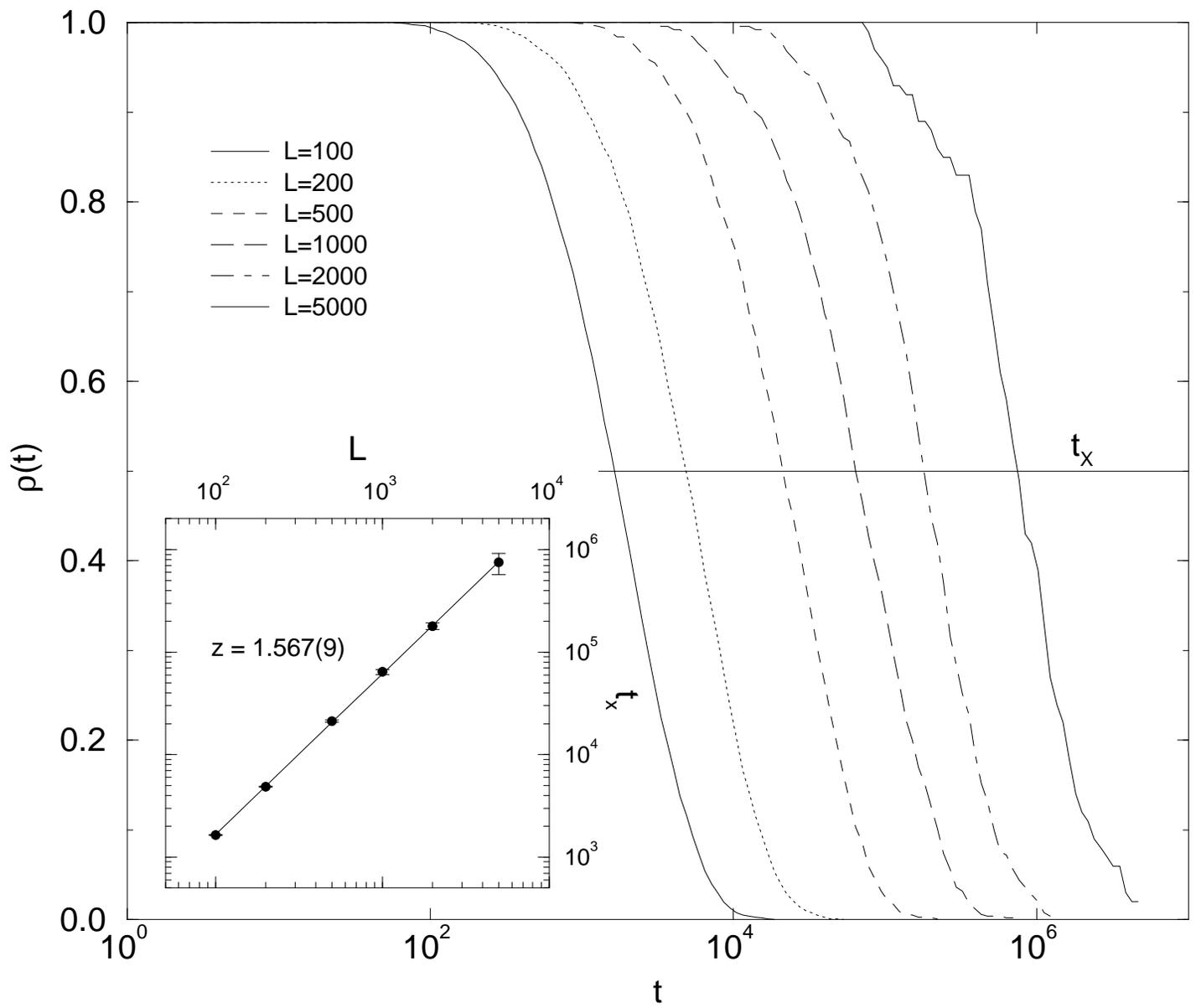
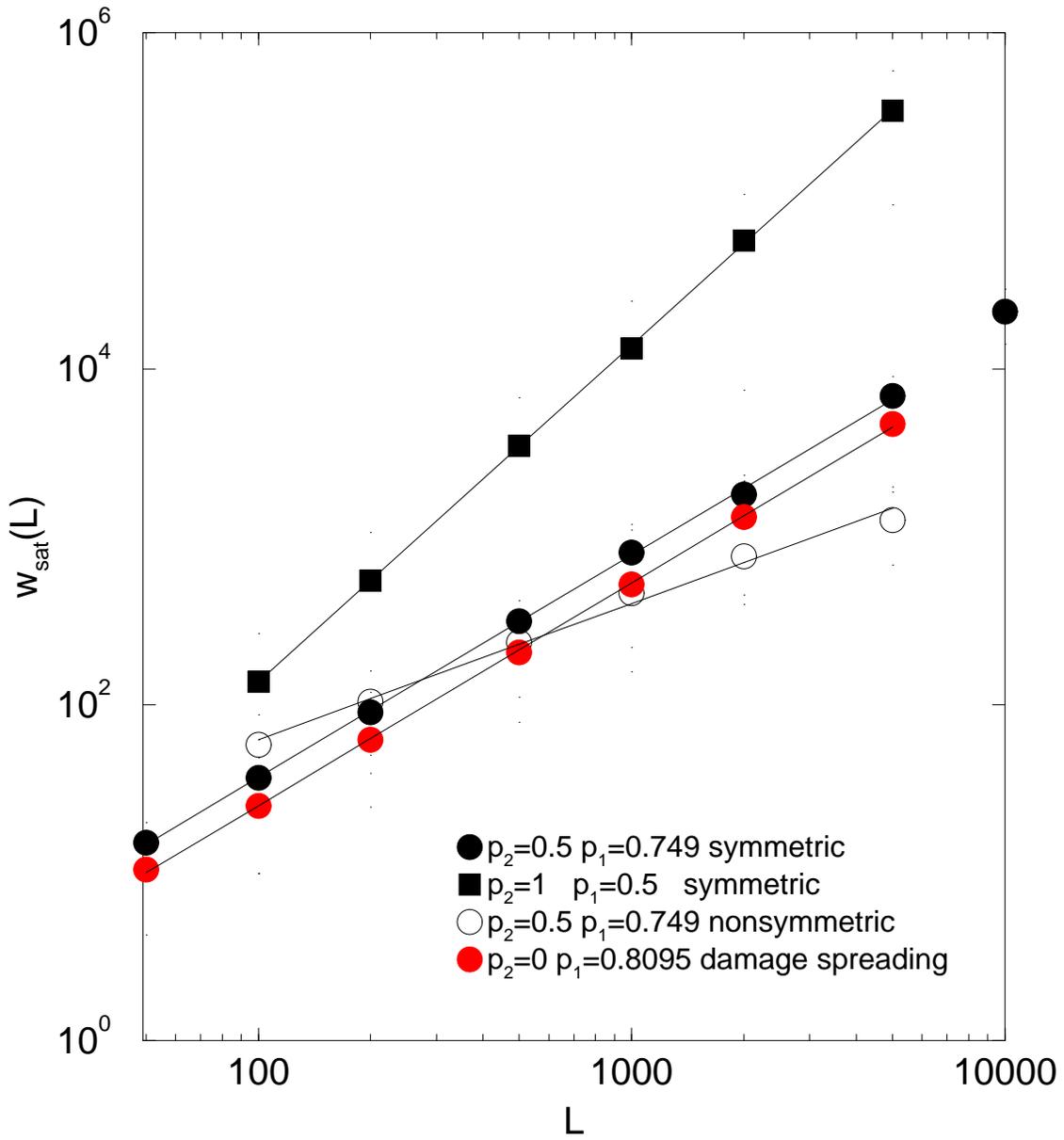
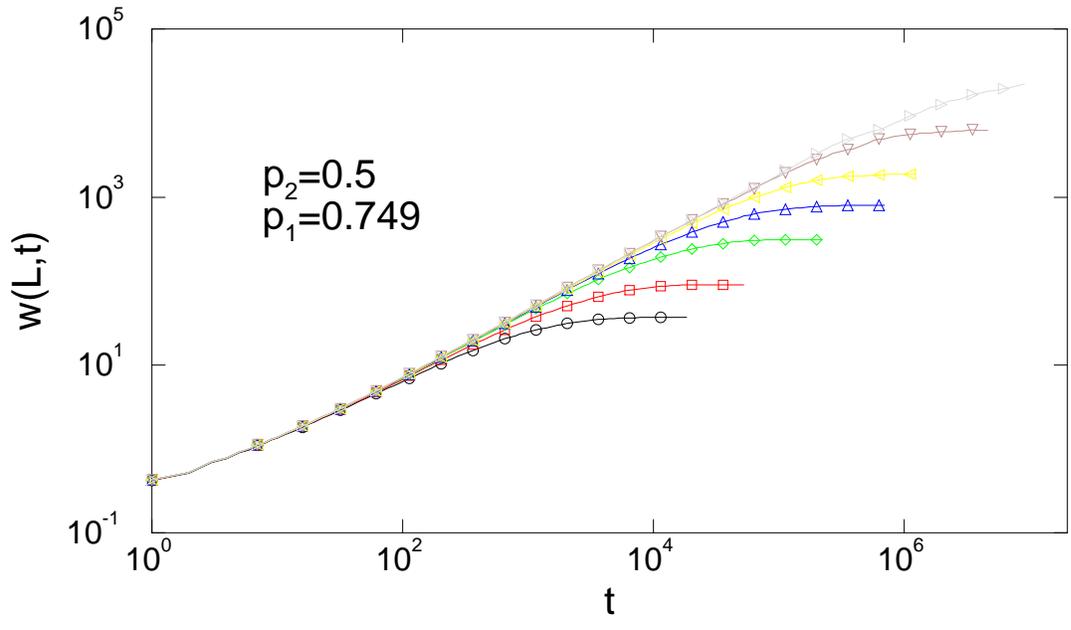
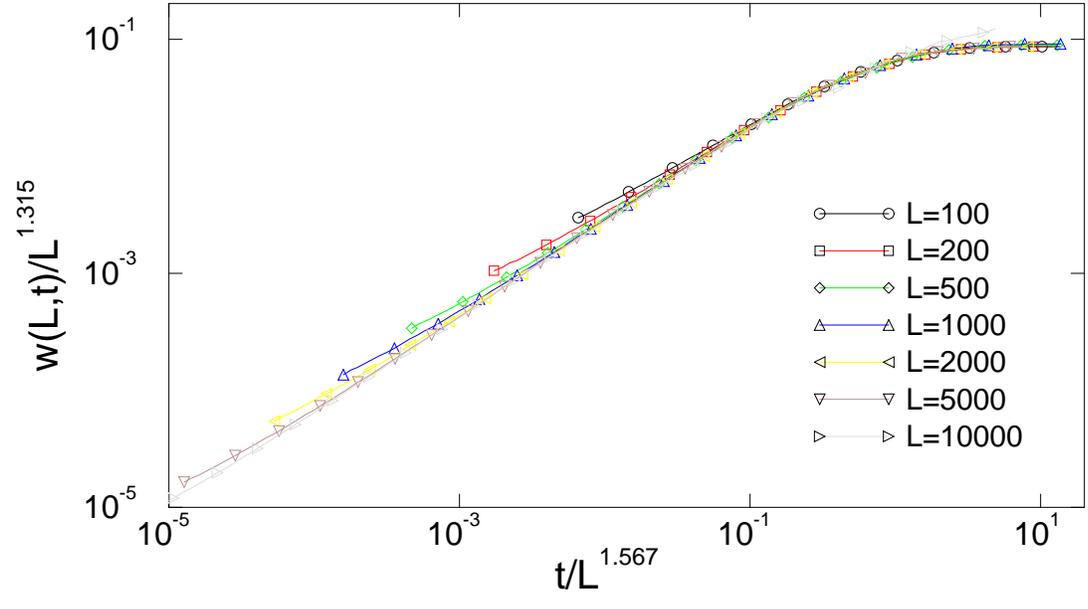
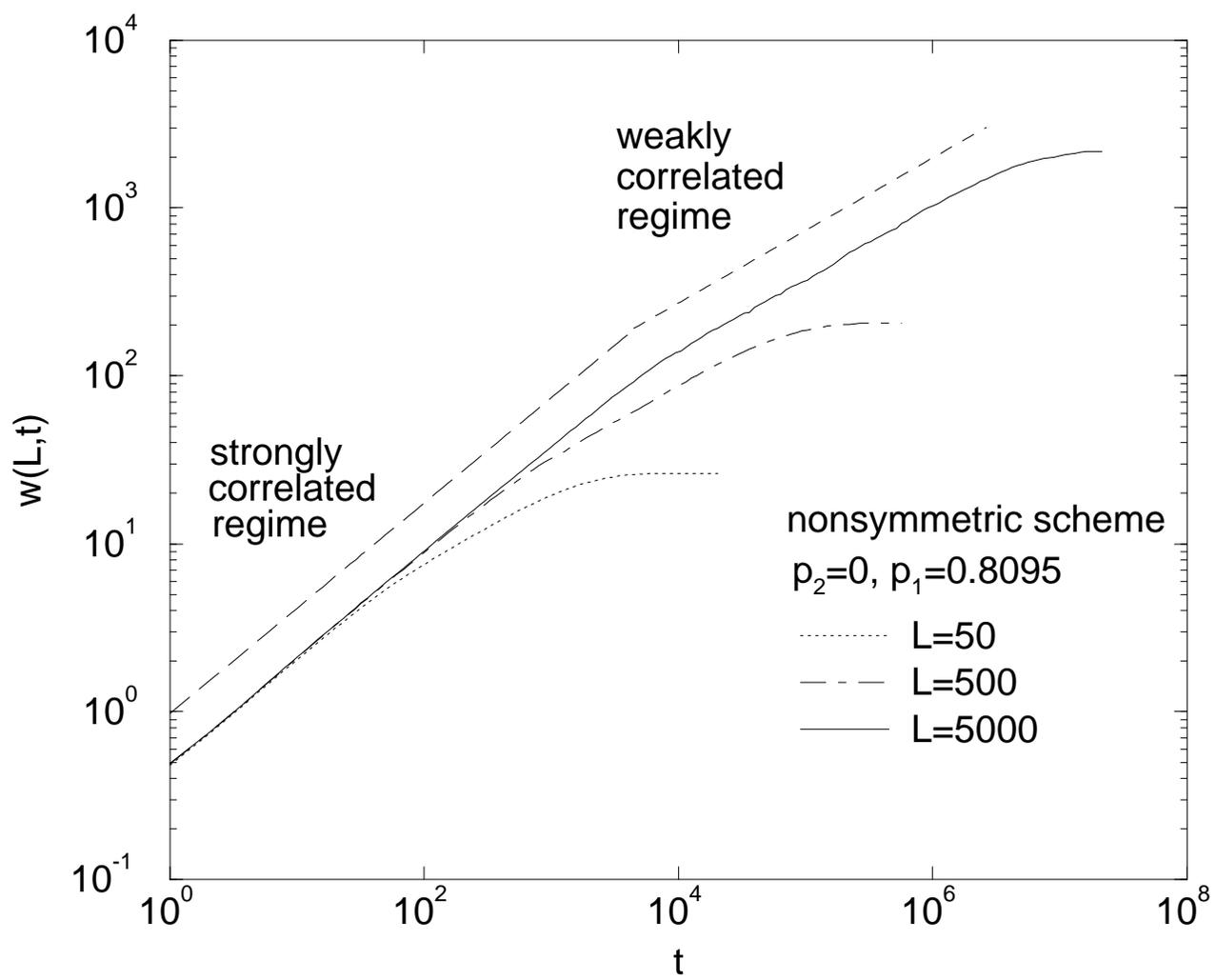
\end{document}